\documentclass[prl,twocolumn,showpacs,preprintnumbers,amsmath,amssymb]{revtex4}

\usepackage{graphicx}
\usepackage{dcolumn}
\usepackage{bm}
\usepackage[]{ams,amsmath,amssymb,epsfig}
\renewcommand{\vec}[1]{\mbox{\boldmath$\mathrm{#1}$}}
\begin{document}
\title{Polarized light bursts from kicked quantum rings}

\author{A.S. Moskalenko}
 \email{andrey.moskalenko@physik.uni-halle.de}
 \altaffiliation[Also at ]{A.F. Ioffe Physico-Technical Institute, 194021 St. Petersburg, Russia}
\author{J. Berakdar}
\affiliation{Institut f\"ur Physik,
Martin-Luther-Universit\"at
 Halle-Wittenberg,  Nanotechnikum-Weinberg, Heinrich-Damerow-St. 4, 06120 Halle,
 Germany}

\date{\today}

\begin{abstract}
 Non-equilibrium quantum rings emit circular polarized subterahertz radiation
 with a polarization degree controllable
 on nano- to picosecond time scales.
 This we conclude using a theory  developed here for the time-dependent
detection of  the circular polarization of polychromatic
radiations, valid for  time scales comparable to the
reciprocal of characteristic emission frequencies. The theory is applied to driven quantum rings whereby
the influence of radiative and non-radiative processes on the properties of the emitted light
is incorporated.
\end{abstract}

\pacs{42.25.Ja,42.65.Re,78.67.--n,41.60.Ap}

\maketitle
 Kekul\'{e}'s work
 on the benzene molecule \cite{Kekule1865}  initiated
  a continuing fascination for  quantum rings. Aharanov-Bohm effect,
  persistent charge and spin currents are just but few examples of the
 intriguing phenomena naturally emerging in these structures  \cite{imry}.
Much efforts are currently  devoted to light-driven
 molecular, nano- and mesoscopic rings  \cite{Kravtsov1993,Chalaev2002,Moskalets2002,Alex_PRB2004,Alex_Europhysics2005,Alex_PRL2005,Moskalenko_PRB2006,Pershin2005,Barth2006b}, a development  fueled
 by the spectacular advance in the controlled fabrication of these materials \cite{mailly,Nitta1999,Lorke2000,Rabaud2001,Fuhrer2001,Yu2007}
 paralleled by equally impressive  progress in laser technology. Light irradiation
  may trigger non-equilibrium charge polarization and ring currents with associated
 ultrafast  electric and magnetic  pulse emission \cite{Kravtsov1993,Alex_PRB2004,Alex_Europhysics2005,Alex_PRL2005,Moskalenko_PRB2006,Pershin2005,Barth2006b,Barth2006a} that can be harnessed for applications, e.g.
     as a polarized light-pulse source
   emitting in a wide range of frequencies. If grown on a tip of  a scanning tunneling microscope (similarly to Ref.~\cite{Schlickum2003}) the ring acts in addition
   as  a local light probe.   In view these prospects a thorough understanding of the time structure, polarization and decay channels of the emitted light is  mandatory. Our purpose here is to provide a general
 quantum mechanical framework  to address these issues.
The  radiation  of electrons moving on a circular orbit was
considered as early as  1907 by Schott in an attempt to explain discrete
atomic spectra using classical electrodynamics \cite{Schott}.
His results, although did not resolve the original problem, were reproduced and became
important much later in the context of  synchrotron radiation
\cite{Schwinger,Landau_Lifshitz_second}.
The polarization characteristics of the angular-resolved
spectrum was   analyzed in Refs.~\cite{Sokolov_Ternov,Ternov_review}. In a synchrotron
specially designed electromagnetic fields  keep the (classical) relativistic electrons on track
 despite energy losses. For quantum rings the confinement potential
forces the  charge to the ring while driven by external fields.
  Unlike the (classical) synchrotron case, for quantum rings
  theory has to deal with non-relativistic, coherent, non-equilibrium charge distributions.
Our results endorse  their possible utilization   as a miniature light sources
with a polarization switchable in picoseconds. \\
{\em Theory.} For a driven quantum ring (with radius $r_0$) we inspect the emitted
far-field radiation at a point $\vec{R},\ R\gg r_0$.
The   far-field  intensity per  solid angle $\Omega=(\theta,\:\phi)$ follows from
 $\frac{{\rm d}I}{{\rm d}\Omega}=\frac{\sqrt{\kappa}}{4\pi c^3}|\langle \hat{\vec{\cal E}}(\vec{R},t)\rangle|^2$ (we use Gau\ss\   units), where $\kappa$ is the medium dielectric constant,
$c$ is the light speed, $\hat{\vec{\cal E}}$ is the electric field operator, and  $\langle ... \rangle$ denotes the expectation value. Extending Refs. \cite{Eberly1977,Raymer1995} to the polarized
case we find the detected time $t$ and frequency $\omega$  dependence of the
intensity $I(\omega,\Omega, t)$  to be
\begin{equation}
\label{Eq:I_alpha}
    \frac{{\rm d}^2 I_\alpha}{{\rm d}\omega{\rm d}\Omega}=\frac{\sqrt{\kappa}}{4\pi
    c^3}\left\langle (\vec{e}_{\alpha}\!\cdot \hat{\vec{\cal
E}}^\dagger)_\mathrm{d}(-\omega,t)\:(\vec{e}_{\alpha}^{\:*}\!\cdot \hat{\vec{\cal
E}})_\mathrm{d}(\omega,t)\right\rangle .
\end{equation}
Here
 $(f)_\mathrm{d}(\omega,t)=\int_{-\infty}^{\infty} f(t') G(t'-t_{\rm d}-t) {\rm
e}^{i\omega t'} {\rm d} t'$,
 $G(t)$ is a
detector function and $t_{\rm d}$ is a  delay time. In terms of the ring's non-equilibrium  electric dipole $\vec{\mu}(t)$, magnetic dipole $\vec{M}(t)$, and
the quadrupole moment $\vec{D}(t)$ (higher terms we neglect) the electric-field coherent part of the emitted light is
 $ \vec{\cal E}(\vec{R},t)=\hat{\vec{n}}\!\times\!\big[\hat{\vec{n}}\times\ddot{\vec{\mu}}(t-t_0)\big]
      \!+\!\sqrt{\kappa}\left\{\hat{\vec{n}}\!\times\!\ddot{\vec{M}}(t-t_0)
      +\frac{1}{6c}\hat{\vec{n}}\!\times\!\big[\hat{\vec{n}}\!\times\!\dddot{\vec{D}}(t-t_0)\big]\right\},$ %
      with
 $t_0=R/c$ and unit vector $\hat{\vec{n}}\parallel \vec{R}$. We use
$    G(t)=\left(\frac{2}{\pi}\right)^{1/4}\!\!\!\frac{1}{\sqrt{\Delta T}}\:{\rm e}^{-t^2/\Delta T^2}$,
 $\Delta T$ is the  local oscillator-pulse duration
\cite{Raymer1995}, and  $t_{\rm d}=t_0$.
Below the following polarization vectors are used: $\vec{e}_{\sigma}$ is   in the plane of
the ring and  perpendicular to  $\hat{\vec{n}}$, $\vec{e}_{\pi}$ is
perpendicular to $\vec{e}_{\sigma}$ and $\hat{\vec{n}}$, $\vec{e}_{\pm
45^\circ}=\frac{1}{\sqrt{2}}(\vec{e}_{\sigma}\pm\vec{e}_{\pi})$, and
$\vec{e}_{\pm}=\frac{1}{\sqrt{2}}(\vec{e}_{\sigma}\pm i \vec{e}_{\pi})$.
Our  aim is to control the
polarization properties of the emitted light by a dynamical control of the rotating charge
polarization of the ring, using half-cycle pulses (HCP) \cite{You1993,Jones1993} or short-laser pulses.
Compared to other methods for  an ultrafast control of
optical fields
\cite{Brixner2001,Brixner2003,Polachek2006,Aeschlimann2007} our
emitted radiation is in the subterahertz range and is controllable
within times comparable to the charge oscillation period in the ring (cf. below).
The light polarization is  classified  by the Stokes
parameters $S_0$, $S_1$, $S_2$, and $S_3$. $S_0$ is the intensity, $S_1$,
$S_2$ and $S_3$ quantify respectively  linearly and circularly polarized light
($P=\sqrt{S_1^2+S_2^2+S_3^2}/S_0$ is the  degree of light polarization).
In term of \eqref{Eq:I_alpha} $S_j$ read
\begin{equation}\label{Eq:S1_omega_t_theta_phi}
     S_{1(2,3)}(\omega,t;\theta,\phi)=
     \frac{{\rm d}^2 I_{\sigma(45^\circ,+)}}{{\rm d}\omega{\rm d}\Omega}-\frac{{\rm d}^2 I_{\pi(-45^\circ,-)}}{{\rm d}\omega{\rm
     d}\Omega}\;.\\
 \end{equation}
 For coherent electric dipole emission we find if
$\hat{\vec{n}}$  is perpendicular to  the ring, i.e. $\theta=0$ ($S^\bot_j\equiv S_j(\theta=0)$)
\begin{equation}\label{Eq:S3_omega_theta_phi_explicit}
     S^\bot_3(\omega,t)=\frac{\sqrt{\kappa}}{4\pi
     c^3}
     2\mathrm{Im}\left[(\ddot{\mu}_y)_\mathrm{d}(-\omega,t)(\ddot{\mu}_x)_\mathrm{d}(\omega,t)\right]\;,
\end{equation}
otherwise $S_3(\omega,t;\theta,\phi)=S^\bot_3(\omega,t)\cos\theta$;
whereas
$     S^\bot_0(\omega,t)=\frac{\sqrt{\kappa}}{4\pi
     c^3}
     \left[\left|(\ddot{\mu}_x)_\mathrm{d}
     (\omega,t)\right|^2+\left|(\ddot{\mu}_y)_\mathrm{d}(\omega,t)\right|^2\right],
$
and the $\phi$-averaged   $S^{\rm }_0(\omega,t,\theta,\phi)$ is given by $S^\bot_0(\omega,t)(1+\cos^2\theta)/2$.
Further we introduce
 $ \bar S_j(t;\theta,\phi)=2\int_0^\infty\!\! \frac{{\rm d}\omega}{2\pi}\;
S_j(\omega,t;\theta,\phi),\ j=0,1,2,3.$
 $\bar S_0(t,\theta,\phi)$ is a time-dependent  power per $\Omega$.
The frequency-integrated circular-polarization degree  is
 $   P_{{\rm circ}}(t;\theta,\phi)= \bar S_3(t;\theta,\phi) /\bar S_0(t;\theta,\phi).$\\
%
%
%
{\em Polarization creation: density-matrix approach.} A realization of ''quantum  nano-synchrotron'' is provided by a spin-degenerate,  thin, isolated (without contacts) two-dimensional
quantum ring with width $d$ ($d\ll r_0$) containing $N$
carriers. The single-particle energies $\varepsilon_{lm}$ and wave functions
$\psi_{lm}(\vec{r})$ are labled  by the radial quantum number $l$ and the angular quantum
number $m$.  For clarity  we assume  $d$ is  small enough such that
 only the lowest radial channel ($l=1$) is populated \footnote{Theory  is extendable to
multiple radial channels as in \cite{Alex_Europhysics2005}.}. In the basis $\psi_{lm}$  the field operator reads $\hat{\Psi}(\vec{r},t)=\sum_m
\hat{a}_m(t)\psi_{1,m}(\vec{r})$ and
the density matrix is given by $\rho_{mm'}(t)=\langle \hat{a}_{m}^{\dag}(t)\hat{a}_{m'}(t)\rangle$
The dipole moment   $\vec{\mu}=\textrm{Tr}[e \vec{r} \rho]$
 ($e$ is the electron charge)
has the components  $\mu_x =2 e r_0 \sum_m \mbox{Re}[\rho_{m\:m-1}]$ and $\mu_y =2 e r_0 \sum_m
\mbox{Im}[\rho_{m\:m-1}]$. The system time scale is given by
the ballistic time $\tau_{_{F}}=2\pi r_0/v_{_F}$, where $v_{_F}$ is the Fermi velocity (typically $\tau_{_{F}}$ is tens of picoseconds).
To trigger  and control the carrier dynamics we utilize
asymmetric monocycle  pulses, so-called  half-cycle pulses (HCPs) \cite{You1993,Jones1993}. For HCPs the pulse duration is chosen as $\tau_d\ll \tau_{_{F}}$.
Application of the pulse at $t=0$ gives the system a kick creating a
time-dependent dipole moment $ \vec{\mu}$  of the ring \cite{Alex_PRB2004,Moskalenko_PRB2006}.
Theoretically, the parameter $\alpha$ is decisive; $\alpha=r_{0}p/\hbar,\ p=-e\int_0^{\tau_d} {\cal E}_{\rm p}(t) \mathrm{d}t$ and ${\cal E}_{\rm p}(t)$ is the electric field strenght of the pulse.
 For a pulse  linearly polarized along the $x$ axis and $\alpha<1$  we find
    $\mu_x(t)\approx-{\alpha} er_0
   \sum_m(f^0_{m-1}-f^0_m) \sin\left[(\varepsilon_m-\varepsilon_{m-1})t/\hbar\right]$,
 where
$f^0_m=\rho_{mm}(t<0)$ is the equilibrium distribution at $t<0$.
 Then at $t=\tau_{_F}/4$ we apply  a second HCP along the $y$-axis creating a rotating dipole (and a charge current \cite{Alex_PRL2005}).

\begin{figure}[t]
  \centering
  \includegraphics[width=6.0cm]{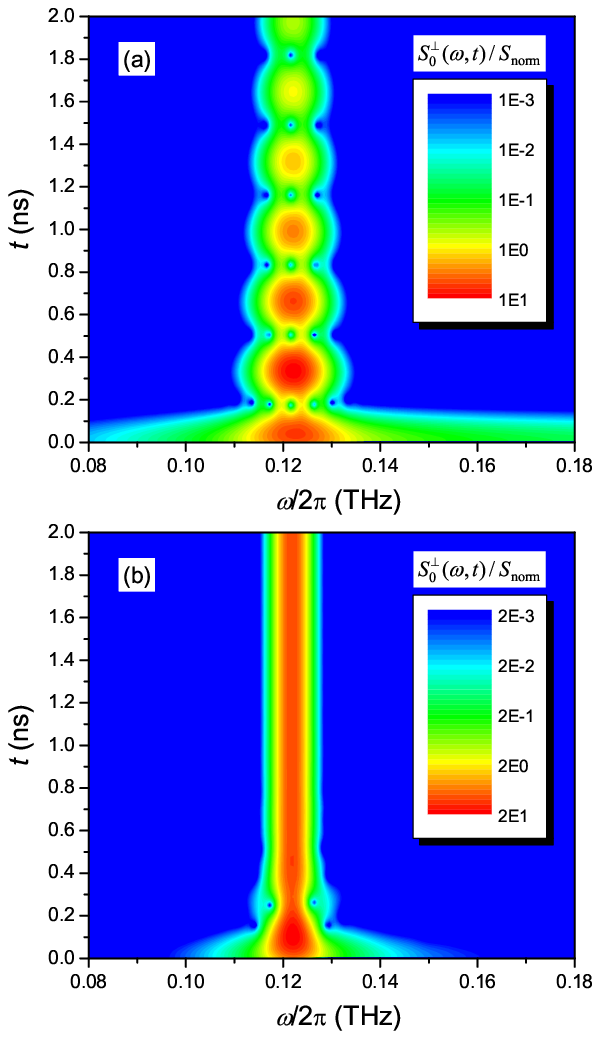}
  \caption{\label{Fig:time_dep_spectrum} (Color online) Time-dependent spectrum of the radiation emitted from a
  quantum ring.  The spectrum
  is normalized to $S_{\rm norm}=\sqrt{\kappa}\frac{e^2r_0^2\omega_{\rm F}^3}{4\pi c^3}$. Parameters
  of the quantum ring are $r_0=0.3~\mu$m, $d=20$~nm, $N=160$. The ring
  is excited (a) by two mutually perpendicular HCPs with kick strength $\alpha=0.4$ and delay time
  $\tau_{_\mathrm{F}}/4$ at $T=4$~K and (b) by a periodic sequence of two mutually perpendicular HCPs with kick strength $\alpha=0.1$,
  delay time $\tau_{_\mathrm{F}}/4$, and period $\tau_{_\mathrm{F}}$, at $T=10$~K. Duration of each HCP is $\tau_d=0.5$~ps, peak field value 33.6~V/cm in (a) and 8.4~V/cm in (b). Detector
  time is $\Delta T=100$~ps.
   }
\end{figure}

Alternatively, short  circular-polarized laser pulses (CPP) with a frequency centered around $\omega=2\pi/\tau_{_{F}}$
create a rotating dipole in the ring. Such pulses or their sequences were suggested previously for exciting rotating polarization and currents in atoms \cite{Barth2007}, molecular rings \cite{Barth2006a,Barth2006b,Nobusada2007}, as well as in nanosize quantum rings and dots \cite{Pershin2005,Rasanen2007}.
For a CPP we can define $\alpha'=\frac{1}{2}r_0p'/\hbar$ with $p'=-e\int_0^{\tau_d} {\cal S}_{\rm p}(t) \mathrm{d}t$, where ${\cal S}_{\rm p}(t)$ is the envelope of the electric field of the CPP.
We have calculated that application of a CPP creates practically the same rotating dipole
moment as in the case of the two $\frac{\tau_{_F}}{4}$-delayed perpendicular HCPs with $\alpha=\alpha'$ for not too strong ($\alpha \gg 1$) pulses \footnote{It is also important that the number of cycles of a CPP is not too large (its frequency distribution is not too narrow) so that all transitions between neighboring levels of the ring in the neighborhood of the Fermi level can take place.}.
 HCPs are insofar favorable as there is no need to fine tune the pulse carrier frequency (as for CPPs);
 also for disordered or deformed rings it is advantageous to use HCPs.
Having created radiating dipoles it is essential for application
to understand how they decay.\\
%
%
%
{\em Radiative decay.}
Radiative damping results from the back action of the
emitted radiation on the non-equilibrium carriers.
Energy loss due to emission causes
relaxation of the density matrix.
Starting from the light-matter
interaction Hamiltonian $\hat{H}_D=-\int \!\!{\rm d}^3r\; \hat{\Psi}^\dagger(\vec{r},t) e\vec{r}\cdot
\hat{\vec{\cal E}}(\vec{r},t) \hat{\Psi}(\vec{r},t)$  the coupled equations
for the dynamics of the field and density operators attain a tractable form if
we allow  for classical (coherent)
electric fields only (semiclassical approximation). Applying the adiabatic (Markov) approximation
\cite{Lehmberg1969,Milonni1975,Khoo1976,Rossi_Kuhn}
we derived the low-temperature decay rate of the dipole moment as
$\gamma=\frac{1}{6}\sqrt{\kappa}\frac{e^2\omega_{_F}^2}{m^*c^3}N$, where
$\omega_{_F}=2\pi/\tau_{_F}$ is the Fermi angular velocity and $m^*$ is the carrier effective mass. For
semiconductor-based  (n-GaAs) ring we have $m^*=0.067m_0$ and $\kappa=12.5$; other typical values are
$r_0=1.35~\mu$m and $N=400$; meaning that $\gamma=1.5\times10^2~{\rm s}^{-1}$. For $r_0=0.3~\mu$m and
$N=160$ we find $\gamma=0.4\times 10^4~{\rm s}^{-1}$. Considering a planar array of $N_r$
identical rings with the array dimension being  smaller than the characteristic
wavelength of the radiation
 the radiative damping rate is found to be
$\gamma_{_\Sigma}=\frac{1}{6}\sqrt{\kappa}\frac{e^2\omega_{_F}^2}{m^*c^3}NN_r$. In practice
rings have a finite spread  $\Delta r_0$ of the ring radius distribution ($\Delta r_0\ll r_0$).
 After a time laps
$\tau_{\rm spr}=\frac{2\pi}{\omega_{_F}}\frac{r_0}{\Delta r_0}$ the electron dynamics in
different rings is out of phase and the intensity of coherent radiation vanishes.
Hence, the
value of $\gamma_{_\Sigma}$ is relevant  only at  short times after the
array excitation, otherwise it provides an upper bound.\\
{\em Spontaneous emission.}
To account for the spontaneous emission contribution to the radiative
damping we study the correlations between the field and density operators
\cite{Rossi_Kuhn} by applying the Hartree-Fock approximation
$\langle\hat{a}_{m_1}^\dagger \hat{a}_{m_2}^\dagger \hat{a}_{m_3} \hat{a}_{m_4} \rangle=\rho_{m_1
m_4}\rho_{m_2 m_3}-\rho_{m_1 m_3}\rho_{m_2 m_4}$ and neglecting correlations involving
two photon operators.
The adiabatic approximation  for the
field-density correlations leads  to  additional contributions to
the decay of the non-diagonal density matrix components $\rho_{mm'}$ due to the spontaneous
emission, namely $\gamma^{\rm sp}_{mm'}=\gamma
\left(\frac{\varepsilon_{m}-\varepsilon_{m'}}{\hbar\omega_{_F}}\right)^3(2+f^0_{m+1}+f^0_{m'+1}-f^0_{m-1}-f^0_{m'-1})$.
Note, the effective spontaneous decoherence rate (decay rate of the dipole moment)
does not exceed $2\gamma$ for low temperatures.\\
{\em Carrier-phonon interaction.}
For weak excitation ($\alpha<1$)  optical phonons are not involved
due to their high energy.  Relaxation of the excited polpulation is however caused by
scattering from longitudinal acoustic phonons. Proceeding as for the
radiative damping and assuming the Debye model for the phonon
dispersion we derived the decay rate of the dipole moment due to the emission of \emph{coherent phonon} waves  as $\gamma^{\rm s}=\frac{1}{\tau_{\rm LA}}F\Big(\frac{\omega_{_F}d}{c_{\rm LA}}\Big)$ if
 $\omega_{_F}<\omega_{_{\rm D}}$, a condition valid for our rings.
Here $\omega_{_{\rm D}}$ is the Debye frequency, $c_{_{\rm LA}}$ is the LA-velocity of sound, the
function $F(y)$ for our quantum-well radial confinement is
$F(y)=8\pi^2y\int_0^{y}\!\!{\rm d}x\; \frac{1}{\sqrt{1-x^2/y^2}}\;
\frac{\sin^2\!(x/2)}{x^2\left[x^2-(2\pi)^2\right]^2}$, and $\frac{1}{\tau_{_{\rm LA}}}=\frac{|D|^2}{\hbar
c_{\rm LA}^2\rho_s d^2 r_0}$, where $D$ is the deformation constant and $\rho_s$ is the lattice
density. For electrons in GaAs  we find $\gamma^{\rm s}=0.8\times 10^6~{\rm s}^{-1}$ for the case
 $r_0=1.35~\mu$m, $d=50$~nm, $N=400$. For $r_0=0.3~\mu$m, $d=20$~nm, $N=160$ we deduce
$\gamma^{\rm s}=2.1\times 10^8~{\rm s}^{-1}$. Note, the wavelength of emitted
phonons is less or on the order of the ring size. Considering thus a
planar ring array the phonon waves emitted from  different places  are out of
phase; a large increase in the damping constant with the number of rings
is thus unlikely.
Density matrix relaxation due to scattering by \emph{incoherent phonons} we
 considered in Ref.~\cite{Moskalenko_PRB2006}, based on which we conclude that the
non-diagonal density matrix components decay due to the spontaneous emission of incoherent phonons
 by the rate $\gamma^{\rm s,sp}_{mm'}= \frac{1}{\tau_{_{\rm
LA}}}\sum_{\nu}(R_m^{m+\nu}+R_{m'}^{m'+\nu})$, where $R^{m}_{m'}= F(q^{m}_{m'}
d)\chi^m_{m'}f^0_{m}$ if $q^{m}_{m'}\in\Big(0,\frac{\omega_{_{\rm D}}}{c_{_{\rm LA}}}\Big)$,
$R^{m}_{m'}= F(q^{m}_{m'} d)\chi^m_{m'}(1-f^0_{m})$ if $q^{m}_{m'}\in\Big(-\frac{\omega_{_{\rm D}}}
{c_{_{\rm LA}}},0\Big)$, and $R^{m}_{m'}=0$ otherwise. Here
$q^{m}_{m'}=(\varepsilon_m-\varepsilon_{m'})/(\hbar c_{_{\rm
  LA}})$, $\chi^m_{m'}=1$ if ${\rm sgn}(m)={\rm sgn}(m')$ and $|q^{m}_{m'}|<\frac{\omega_{_{\rm D}}}{c_{_{\rm
LA}}}$, $\chi^m_{m'}=0$ otherwise.
Using the expression for $\gamma^{\rm s,sp}_{mm'}$ we
calculated numerically the effective decay rate of the dipole moment for different parameters of
the quantum ring. For the above mentioned  ring parameters the decay rate
rises from  $10^{8}$~s$^{-1}$ at $T=1$~K to  $10^{10}$~s$^{-1}$ at $T=10$~K (see also
Ref.~\cite{Moskalenko_PRB2006}). Thus,  scattering by  incoherent phonons is the main
source for our dipole moment relaxation. \\
\emph{Numerical results.}
For illustration we focus on a quantum ring excited by two mutually perpendicular HCPs delayed by
$\tau_{_\mathrm{F}}/4$. Fig.~\ref{Fig:time_dep_spectrum} shows the  time-dependent
spectrum of the radiation emitted perpendicular to the ring plane  ($\theta=0$).
Only the electric dipole contributes to the
radiation spectrum in this geometry.
Calculating the time-dependent spectrum we use the detector time $\Delta T=100$~ps.
The numerically calulated  time-dependent spectrum Fig.~\ref{Fig:time_dep_spectrum} exihbits
 repeated  light bursts centered at approximately
$\omega_{_F}$ and with decaying peak values due to the  relaxation pathways discussed above.
 The bursts correspond to revivals of the charge polarization dynamics \cite{Moskalenko_PRB2006}.
The abrupt radiation switch-on   upon the  charge-polarization generation  results
in a relatively broad spectrum at short times.
Particularly  interesting is the degree of  circular polarization at $\theta=0$,  which we denote as
$P_\mathrm{circ}^\bot(t)$.
 For a detected frequency range
$[0.5\omega_{_\mathrm{F}},1.5\omega_{_\mathrm{F}}]$
we find  $P_\mathrm{circ}^\bot(t)> 0.99$. Driving the
rotating dipole in sequence   effectively stabilizes the time-dependent spectrum
 reaching a practically stationary state (cf. Fig.~\ref{Fig:time_dep_spectrum}b). In this case
 $P_\mathrm{circ}^\bot(t)> 0.999$ is achievable for the mentioned frequency range.

\begin{figure}[t]
  \centering
  \includegraphics[width=5.5cm]{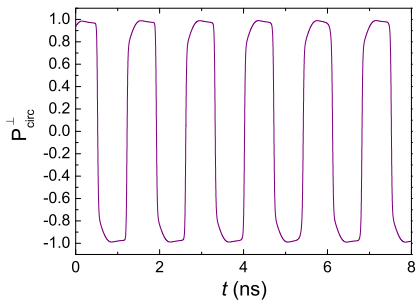}
  \caption{\label{Fig:figure_A2} (Color online) Dynamics of the detected circular
  polarization.  The ring parameters
  are $r_0=1.35~\mu$m, $d=50$~nm, $N=400$, and $T=4$~K.
  For the parameters of the excitation see  the text.
  Detector time is $\Delta T=100$~ps.}
\end{figure}

Having demonstrated the possibility of an ultrafast generation of  circular polarized
subterahertz light we are interested in the ultrafast control of  the degree of the circular
polarization. For illustrating the ubiquitous nature of the effect we
choose a different radius of 1.35~$\mu$m.  The pulses parameters are $\alpha=0.2$,  $\tau_d=3$~ps and a time delay
 $\tau_{_\mathrm{F}}/4$. Furthermore, for a sequence of pulses we assume a period of  0.7 ns.
 Each applied pulse sequence changes the sense of rotation of the ring dipole moment. This
leads to a sign change of  $P_\mathrm{circ}^\bot$, as demonstrated
 in Fig.~\ref{Fig:figure_A2}; i.e.
the photons are emitted in portions
with an alternating helicity. Generally,  more complicated pulse sequences   allow the control of
the chirality of the each emitted photon portion.

Summarizing, we developed a  theory for the  ultrafast, time-dependent detection of the light polarization properties and applied it to the emission  from quantum rings. As an illustration we presented calculations for the time-dependent spectrum accounting for relaxation effects. Applying an appropriate sequences of the half-cycle pulses (or short circular polarized laser pulses) allow to generate circular polarized subterahertz light on picosecond time scales and to change
and control the degree of the circular polarization of the emitted photons in subnanoseconds. For a sufficient intensity of emitted light one can either use an
array of rings \cite{mailly,Yu2007,Kleemans2007} or excite a single ring far beyond the weak excitation regime considered in the
present paper. Our treatment is directly applicable to the first case whereby it is necessary to
provide a small dispersion of  the ring sizes in the array. The proposed theory is extendable to  higher excitations, to quantum dots and quantum rings with an impurity
\cite{Zipper2006} as well as to graphene rings \cite{Recher2007}   and  chiral and non-chiral molecular rings
\cite{Kanno2006,Kanno2007,Barth2006a,Barth2006b,Nobusada2007} and to atoms \cite{Barth2007}. Finally, the developed theory is applicable to the time-dependent detection of the circular polarization of quantum light \cite{Walmsley2008,Shields2007}.

\end{document}